\documentclass[twocolumn,superscriptaddress]{revtex4-1}

\usepackage{amssymb}
\usepackage{amsmath}
\usepackage{bbold}
\usepackage{color}
\usepackage[usenames,dvipsnames]{xcolor}
\usepackage{graphicx}
\usepackage{commath}

\graphicspath{{figura_k2/}{figura_k3/}{figura_sketch/}}

\begin{document}

\title{Counting the learnable functions of structured data}

\author{Pietro Rotondo}
\affiliation{School of Physics and Astronomy, University of Nottingham, Nottingham, NG7 2RD, UK}
\affiliation{Centre for the Mathematics and Theoretical Physics of Quantum Non-equilibrium Systems, University of Nottingham, Nottingham NG7 2RD, UK}
\author{Marco Cosentino Lagomarsino}
\affiliation{Universit\`a degli Studi di Milano, Via Celoria 16, 20133 Milano, Italy}
\affiliation{I.N.F.N.~Sezione di Milano}
\author{Marco Gherardi}
\email[Corresponding author: ]{marco.gherardi@mi.infn.it}
\affiliation{Universit\`a degli Studi di Milano, Via Celoria 16, 20133 Milano, Italy}
\affiliation{I.N.F.N.~Sezione di Milano}

\begin{abstract}
  Cover's function counting theorem is a milestone in the theory
  of artificial neural networks. It provides an answer to the
  fundamental question of determining how many binary assignments
  (dichotomies) of $p$ points in $n$ dimensions can be linearly
  realized.  Regrettably, it has proved hard to extend the same approach
  to more advanced problems than the classification of points.
  In particular, an emerging necessity is to find methods
  to deal with structured data, and specifically with non-pointlike patterns. 
  A prominent case is that of invariant recognition,
  whereby identification of a stimulus is insensitive to irrelevant transformations
  on the inputs (such as rotations or changes in perspective in an image).
  An object is therefore represented by an extended
  perceptual manifold, consisting of inputs that are classified similarly.
  Here, we develop a function counting theory for structured data of
  this kind, by extending Cover's combinatorial technique, and we derive analytical
  expressions for the average number of dichotomies of generically correlated
  sets of patterns. 
  As an application, we obtain a closed formula for the capacity of a
  binary classifier trained to distinguish general polytopes of any dimension.
  These results may help extend our theoretical understanding of
  generalization, feature extraction, and invariant object recognition
  by neural networks.
\end{abstract}

\maketitle

\section{Introduction}
Machine learning and deep learning demonstrate astonishing results in
applications \cite{Hinton2012,Goodfellow2014,Goodfellow2016}, sometimes beyond our theoretical reach. This provides a
formidable challenge for theorists who wish to develop a framework for
their understanding
\cite{BaldassiBorgs:2016,Baity2018}.
A landmark achievement in learning theory is Cover's
function counting theorem, which counts the number of binary
classification functions, or ``dichotomies'', that can be realized by
given architectures~\cite{cover1965}.
This foundational result allowed to quantify the complexity of a
learning model and the advantage gained in using non-linear kernels,
provided a benchmark for the performance of both artificial and
natural neural networks,
and is a handy tool for several applications
\cite{BrunelHakim2004,EngelVanDenBroeck:BOOK,HertzKroghPalmer:BOOK,Opper1996,
GanguliSompolinsky:2010,ChungLeeSompolinsky:2018}.
Other commonly used methods in this area
come from statistical physics
(pioneered by E.~Gardner
\cite{Gardner:1987,GardnerDerrida:1988}).
With respect to these, Cover's method has the
advantage of offering a simple geometric insight and of being valid at
finite number of dimensions, while statistical physics methods
typically apply in the ``thermodynamic limit'' of infinite dimensions.
Yet, despite its benefits and relative simplicity, Cover's analytical
technique has so far eluded efforts to extend it
\cite{EngelVanDenBroeck:BOOK}.

Uncorrelated random patterns are commonly taken as a simplifying
assumption for the theoretical investigation of artificial neural networks. 
Yet, it is becoming apparent that providing a theoretical
framework that includes structure in the input data is essential. This
need is emerging in different contexts: 
(a) The invariant
representation of perceptual stimuli by brains (e.g., the coherent
perception of differently rotated and rescaled objects in vision,
or the recognition of the same sound in different acoustic environments
in audition)
prompted the formalization of perceptual manifolds as extended patterns 
\cite{Tenenbaum:2000,Seung:2000,Roweis:2000,Ranzato:2007,GoodfellowLee:2009,
AnselmiLeibo:2016,
ChungLeeSompolinsky:2016,ChungLeeSompolinsky:2018,ChungCohen:2018}.
Perceptual manifolds are the regions
in input space corresponding to all variations of a stimulus that do not
modify the object's identification.
(b) The discovery of spatial maps in rodent brains
\cite{okeefe:1971}
motivated extensions of associative memory models to attractors
that are not point-like but occupy a region in configuration space
\cite{CoccoMonasson:2018}.
(c) The problem of local generalization and robustness to noise,
a main theme of machine learning,
can be cast as a problem of non-pointlike patterns
\cite{SzegedyZaremba:2013,Novak:2018,Borra:2019}.
(d) The description of the input patterns as modular combinations of
elementary features
(a well studied aspect of empirical datasets \cite{PangMaslov:2013,Mazzolini:2018}),
was shown to induce a multi-layer structure in certain network
architectures \cite{Mezard:2017}.

Here, we develop a theory that extends Cover's approach to non
point-like patterns, by counting only those dichotomies
that assign the same label to different variants of the same input.
Our theory
(i) enables the exact computation of the
(average) number of dichotomies of structured data, (ii) gives direct
access to quantities at finite size, and (iii) naturally disentangles
combinatorial and geometric aspects, thus lending itself to further
generalizations.

\section{Number of admissible dichotomies}
\begin{figure}
\includegraphics[scale=1]{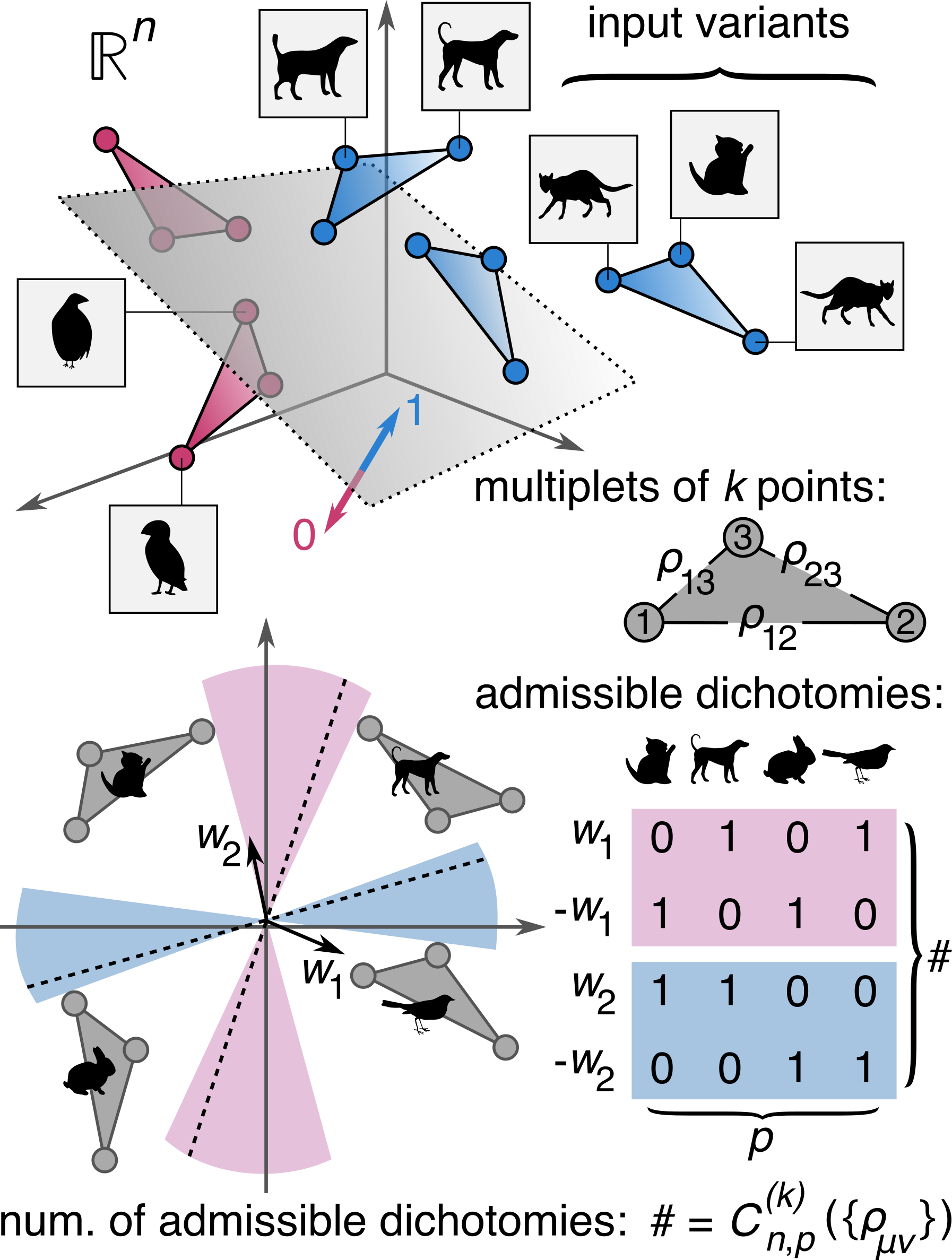}
\caption{
(Top) A dichotomy is identified by a hyperplane (in grey),
separating differently labeled data points,
(e.g., mammals from birds, mapped respectively to 1 and 0).
Data are structured in multiplets of $k$ input variants (here $k=3$).
Each input variant is a point in $\mathbb{R}^n$,
and each multiplet is characterized by the $k(k-1)/2$ overlaps
between its points
(here $\rho_{12}$, $\rho_{23}$, and $\rho_{13}$).
A dichotomy is admissible only
if it is constant on each multiplet, i.e.,
if the separating hyperplane does not intersect any polytope (triangles here).
(Bottom) Given a structured data set
(here $p=4$ multiplets of $k=3$ points in $n=2$ dimensions),
we count the number $C_{n,p}^{(k)}$ of
admissible dichotomies.
}
\label{fig:sketch}
\end{figure}

The central quantity obtained by Cover's function counting method is the number $C_{n,p}$ of
linearly-realizable dichotomies of $p$ points $\xi_1,\ldots,\xi_p$ in $n$ dimensions.
A dichotomy of this set
is a function $\phi$
mapping each point $\xi_i$ to its $\left\{0,1\right\}$ binary label
(see Fig.~\ref{fig:sketch}).
A linearly-realizable dichotomy is identified by a vector $w\in\mathbb{R}^n$:
\begin{equation}
\phi\left(\xi_i\right)=\theta\left(\xi_i\cdot w\right),
\end{equation}
where $\theta$ is the Heaviside theta function.
The hyperplane perpendicular to the vector $w$ separates the
space into two half-spaces, where the points mapped to $0$ and $1$ lie respectively.
There are $2^p$ dichotomies, but only $C_{n,p}$ of them are linearly realizable.
We focus on linearly realizable dichotomies, and will therefore omit 
this specification when it is clear from the context.

%
It turns out that $C_{n,p}$
does not depend on the $\xi_i$'s, as long as they are
in general position
(meaning that no subset of $n$ points is linearly dependent)
\cite{cover1965}.
Structure in the data may thus appear not to affect $C_{n,p}$ at all.
However, in general we do not wish to admit all possible dichotomies.
For instance, among the hand-written digits in MNIST we could choose
to admit dichotomies separating ``1'' and ``I'', but not two similar-looking ``0''s.
Our definition of structure is based on such a restriction:
a data set is qualified as structured
whenever only a subset of all possible dichotomies 
is considered admissible.
$C_{n,p}$ will then be the number of admissible dichotomies that can be realized linearly.

Here we focus on a rather general definition of admissibility,
inspired by the literature cited above.
We consider datasets of $kp$ points,
structured as $p$ multiplets of $k$ points each.
A dichotomy $\phi$ is admissible if different points $\xi$ in the same multiplet
are classified coherently, i.e., if $\phi(\xi)$ is constant on each multiplet.
We will restrict the points $\xi$ to lie on the unit sphere $S^{n-1}$,
meaning that $\xi^2=1$, but this technical requirement can be easily relaxed.
(A useful consequence of this is that setting the overlap between two points determines
their distance.)
The ensemble we consider fixes all the overlaps between the points
in a multiplet, equally for all multiplets,
but the relative positions and orientations of the multiplets are unspecified.
The quantities we will compute are averages over all possible
positions and orientations of the multiplets.

Because of the convexity of linear separability,
separating the multiplets is equivalent to separating
the polytopes whose vertices are the points in the multiplets.
(These polytopes play the role of the perceptual manifolds of Ref.~\cite{ChungLeeSompolinsky:2018}.)
For instance, $k=2$ corresponds to segments, $k=3$ to triangles,
$k=4$ to tetrahedra.

\section{Single points ($k=1$)}
Let us first outline Cover's original computation.
Imagine starting with $p$ points and adding the $(p+1)$th point 
$\xi_{p+1}$ to $\left\{\xi_1,\ldots,\xi_p\right\}$.
For each dichotomy $\phi$ of the $p$ points $\xi_1,\ldots\xi_p$
one of two possibilities is satisfied: either (i)
$\phi$ can be realized by a hyperplane passing through $\xi_{p+1}$
(equivalently, $\phi$ can be realized by a vector $w$ such that $\xi_{p+1}\cdot w=0$),
or (ii) it can not.
If (i) is true, then
$w$ can be rotated infinitesimally to yield both $\xi_{p+1}\cdot w\gtrless 0$;
otherwise, the half-space where $\xi_{p+1}$ lies is fixed.
Therefore, for each dichotomy $\phi$ of $\left\{\xi_1,\ldots,\xi_p\right\}$ satisfying (i)
there are 2 different dichotomies $\phi_1$ and $\phi_2$ of $\left\{\xi_1,\ldots,\xi_p,\xi_{p+1}\right\}$
agreeing with $\phi$ on the common points
[i.e., such that $\phi_{1,2}(\xi_i)=\phi(\xi_i)$ for $i=1,\ldots,p$].
If the number of dichotomies satisfying (i) is $M$, then
the number of those satisfying (ii) is $C_{n,p}-M$,
and one can write $C_{n,p+1}=2M+C_{n,p}-M$.
The condition (i) is in the form of a single linear constraint,
therefore $M$ is the number of dichotomies of $p$ points in
$n-1$ dimensions, $M=C_{n-1,p}$.
Thus $C_{n,p}$ satisfies the recursion
\begin{equation}
\label{eq:cover_recursion}
C_{n,p+1}=C_{n,p}+C_{n-1,p},
\end{equation}
with boundary conditions $C_{n>0,1}=2$ (a single point can be classified either way) and $C_{0,p}=0$.

The solution to Eq.~(\ref{eq:cover_recursion}) can be obtained
by observing that the contribution of the boundary value
$C_{n-i,1}$ to $C_{n,p}$
is given by the number of directed paths $\{\gamma_j\}_{j=1,\ldots,p}$,
with $\gamma_j\in\mathbb{N}$,
that start from $\gamma_1=n-i$ and end in $\gamma_p=n$, 
where at each step $\gamma_{j+1}$ can be either 
$\gamma_j$ or $\gamma_j+1$.
The number of such paths is simply the binomial coefficient
${{p-1}\choose i}$.
Summing over the boundary gives
\begin{equation}
\label{eq:cover_solution}
C_{n,p}=2\sum_{i=0}^{n-1} {{p-1} \choose i},
\end{equation}
where it is assumed that ${{p-1}\choose i}=0$
whenever $i>p-1$.

Let us consider the fraction $c_{n,p}$ of linearly realizable dichotomies
$c_{n,p} = C_{n,p}/2^p$.
For finite $n$ and $p$, the capacity $\alpha_\mathrm{c}$ can be defined as the ratio
$p/n$ at which half of all dichotomies can be realized:
$c_{n, n \alpha_\mathrm{c}} = 1/2$.
From the explicit expression (\ref{eq:cover_solution}) one sees that
$c_{n,p}=1$ if $p\leq n$, 
$c_{n,p}\to 0$ for $p\to\infty$,
and $c_{n,2n}=1/2$,
which pinpoints the well-known capacity $\alpha_\mathrm{c}=2$.

\section{Segments (doublets, $k=2$)}
The first step towards the general problem is the case where data are structured
as pairs of points.
Alongside the set of points $\xi=\left\{\xi_1,\ldots,\xi_p\right\}$,
let us consider another set $\bar\xi=\left\{\bar\xi_1,\ldots,\bar\xi_p\right\}$.
The multiplets discussed above are the doublets $\{\xi_i,\bar\xi_i\}$.
Each doublet is such that the overlap
between the two partners is fixed:
\begin{equation}
(-1,1)\ni\rho = \frac{1}{n} \xi_i\cdot\bar\xi_i
\end{equation}
for all $i$.
The admissible dichotomies $\phi$ are those for which
$\phi(\xi_i)=\phi(\bar\xi_i)$ for all $i$;
their total number is $2^p$.

The recursion step now corresponds to the addition
of the $(p+1)$th doublet $\{\xi_{p+1}$, $\bar\xi_{p+1}\}$.
Repeating Cover's reasoning for the point $\bar\xi_{p+1}$ alone gives
a number of dichotomies equal to $Q_{n,p}=C_{n,p}+C_{n-1,p}$.
This is the number of dichotomies of the set
$\{\xi_1,\bar\xi_1,\xi_2,\bar\xi_2,\ldots,\xi_p,\bar\xi_p,\bar\xi_{p+1}\}$
that are admissible on the first
$p$ doublets [meaning that $\phi(\xi_i)=\phi(\bar\xi_i)$ for all $i=1,\ldots,p$].
A number $R_{n,p}$ of such dichotomies are realizable by a hyperplane
passing through the point $\xi_{p+1}$.
These are all admissible, thanks to the freedom
in the choice of $\phi(\xi_{p+1})$
by an infinitesimal adjustment of the hyperplane.
Among the other $Q_{n,p}-R_{n,p}$ dichotomies, on average, a fraction $\Psi_2$
will happen to assign the same label to $\xi_{p+1}$ and $\bar\xi_{p+1}$.
$\Psi_2$ can be computed as the fraction of hyperplanes
keeping $\xi_{p+1}$ and $\bar\xi_{p+1}$
in the same half-space; the calculation is carried out in the Appendix.
Importantly, $\Psi_2$ is a function of the overlap $\rho$ alone:
\begin{equation} \label{eq:Psi2}
\Psi_2(\rho)=\frac{2}{\pi}\arctan\sqrt{\frac{1+\rho}{1-\rho}}.
\end{equation}
Note that $\Psi_2(\rho)=1-\Psi_2(-\rho)$ as expected from its definition.
The foregoing argument brings to
estimate the total number of admissible dichotomies as
\begin{equation}
C_{n,p+1}=\Psi_2(\rho)(C_{n,p}+C_{n-1,p})+\left[1-\Psi_2(\rho)\right]R_{n,p}.
\end{equation} 
In order to compute $R_{n,p}$ it suffices to repeat Cover's
reasoning with respect to the point $\bar\xi_{p+1}$, 
this time in $n-1$ dimensions because of the constraint imposed
by the hyperplane passing through $\xi_{p+1}$, thereby
obtaining 
\begin{equation}
R_{n,p}=C_{n-1,p}+C_{n-2,p}.
\end{equation}
Finally the recursion for $C_{n,p}$ reads
\begin{equation}
\label{eq:the_recursion_k2}
C_{n,p+1} = 
\Psi_2(\rho) C_{n,p} + C_{n-1,p} + \left[1-\Psi_2(\rho)\right] C_{n-2,p}.
\end{equation}

The boundary conditions are now slightly different than those for
the case $k=1$ in Eq.~(\ref{eq:cover_recursion}).
In fact, in $n=1$ dimension the number of admissible dichotomies
of a single pair of points ($p=1$) is $2$ only when both
points lie on the same half-line, otherwise it is $0$;
on average, it is $2\Psi_2(\rho)$.
The boundary conditions are then
\begin{equation}
\begin{split}
C_{0,p}&=0,\\
C_{n>0,1}&=2\left\{1-[1-\Psi_2(\rho)]\delta_{n,1}\right\}.
\end{split}
\end{equation}

To find the solution of the recursion (\ref{eq:the_recursion_k2}),
similarly to the single point case,
consider all the directed paths $\{\gamma_j\}_{j=1,\ldots,p}$
propagating from the boundary to $C_{n,p}$,
where $\gamma_{j+1}$ at each step can be
$\gamma_j$, $\gamma_j+1$, or $\gamma_j+2$.
Contrary to the one point case, different paths
with the same endpoints can now give
different contributions to $C_{n,p}$, since
the three types of steps correspond to three different factors
($\Psi_2$, $1$, and $1-\Psi_2$ respectively).
The contribution $K_{i,p}$ of a path
from $\gamma_1=n-i$ to $\gamma_p=n$ is
\begin{equation}
K_{i,p}= \sum_{m=0}^{p-1}
{{p-1} \choose {m, i-2m}}
\Psi_2(\rho)^{p-1-i+m} \left[1-\Psi_2(\rho)\right]^m,
\end{equation}
%
where the multinomial coefficient is defined as
\begin{equation}
{n \choose {m_1, m_2}} = \frac{n!}{m_1! m_2! \left(n-m_1-m_2\right)!}
\end{equation}
(with the obvious analytical extension for negative factorials).
Summation over the non-zero boundary $i=0,\ldots,n-1$ yields
the number of admissible dichotomies
\begin{equation}
\label{eq:solution_k2}
C_{n,p} = 2 \sum_{i=0}^{n-2} K_{i,p} + 2\Psi_2(\rho)K_{n-1,p}.
\end{equation}
%
It is easy to see (by the multinomial theorem) that
$C_{n,p}=2^p$ if $p\leq n/2$;
this locates the usual Vapnik-Chervonenkis dimension \cite{VapnikChervonenkis:1971},
$d_\mathrm{VC}=n$, as the total number of points is $2p$.
An estimate for the capacity, valid for large $n$,
can be obtained by approximating Eq.~(\ref{eq:solution_k2}) as
\begin{equation}
C_{n,p}\approx 2\sum_{i=0}^{n-1} K_{i,p}.
\end{equation}
The capacity $\alpha_\mathrm{c}$ is such that
\begin{equation}
C_{p/\alpha_\mathrm{c},p}\approx 2^{p-1},
\end{equation}
i.e., it corresponds to
the value of $n$ for which the sum of $K_{i,p}$ takes
half its maximum value.
The quantity $K_{i,p}$ can be interpreted
as the partition function
of an ensemble of directed random walks $\{\gamma_j\}_{j=1,\ldots,p}$
of $p-1$ steps,
with the same boundary conditions as for $k=1$,
and the following transition probabilities:
$P\left(\gamma_j \to \gamma_j\right) = \Psi_2/2$,
$P\left(\gamma_j \to \gamma_j+1\right) = 1/2$,
$P\left(\gamma_j \to \gamma_j+2\right) = (1-\Psi_2)/2$.
The normalization factor $2$ at the denominator is the sum of the weights
$\Psi_2$, $1$, and $1-\Psi_2$.
The capacity therefore corresponds to
the median of the distribution function of the walk's endpoint $i$.
We approximate the median with the mean
%
\begin{equation}\label{eq:imean}
\bar\imath = (p-1) \sum_{l=0}^2 l P\left(\gamma_j \to \gamma_j + l \right),
\end{equation}
which evaluates to
$\bar\imath = ( 3/2 - \Psi_2) (p-1)$, and finally we obtain
\begin{equation}
\label{eq:capacity_k2}
\alpha_\mathrm{c} \approx \frac{p-1}{\bar\imath} 
= \frac{2}{3-2\Psi_2(\rho)}.
\end{equation}
This result, with $\Psi_2$ given by Eq.~(\ref{eq:Psi2}), 
was found in \cite{LopezSchroder:1995} by means of replica calculations,
and appeared more recently in other contexts in 
\cite{Borra:2019,ChungLeeSompolinsky:2016}.
Our derivation is somewhat more elementary,
and naturally highlights the role of the geometric quantity $\Psi_2(\rho)$.

\begin{figure}
\includegraphics[scale=1.1]{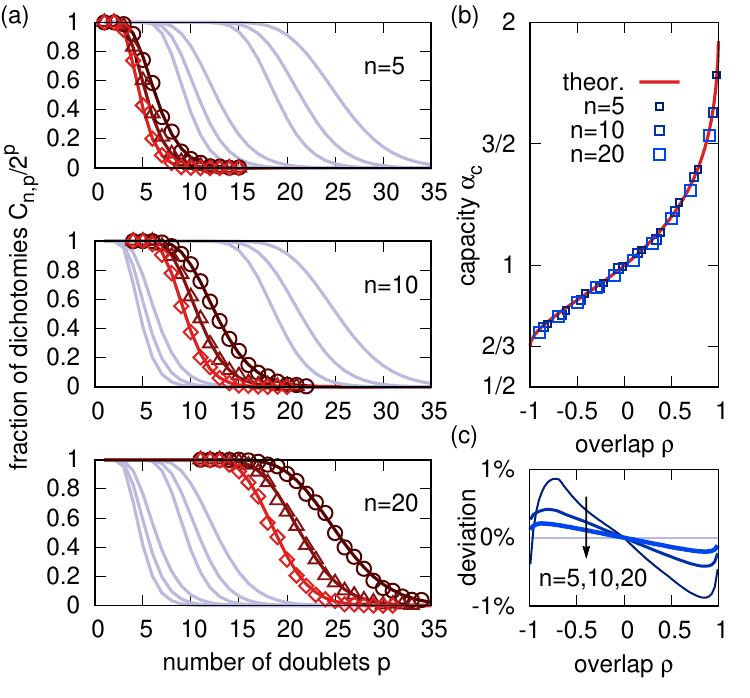}
\caption{
Theory (solid lines) versus numerical results for $k=2$, obtained
by training a linear classifier with the perceptron algorithm.
(a) Fraction of admissible dichotomies (y axis)
as a function of number of doublets (x axis)
in dimensions $n=5,10,20$ for different values of the overlap
$\rho=0.6\;(\circ)$, ${0.2\;(\vartriangle)}$, ${-0.2\;(\diamond)}$.
The theoretical curves are given by Eq.~(\ref{eq:solution_k2});
grey lines are just for comparing the three values of $n$
on the same graph.
(b) The capacity [Eq.~(\ref{eq:capacity_k2})] (y axis) as a function of
the overlap $\rho$ (x axis).
(c) Finite-size deviation of the capacity
[obtained by solving numerically $C_{n,\alpha_\mathrm{c}n}=2^{\alpha_\mathrm{c}n-1}$ 
at fixed $n$ with $C_{n,p}$
given by Eq.~(\ref{eq:solution_k2})]
from the large-$n$ prediction Eq.~(\ref{eq:capacity_k2}).
[Each point in (a) is a fraction over 1000 independent trials;
the capacity in (b) is obtained by linearly interpolating data such
as those in (a).]
}
\label{fig:k2}
\end{figure}

Figure~\ref{fig:k2} compares the analytical formulas
(\ref{eq:solution_k2}) and (\ref{eq:capacity_k2}) with numerical results obtained
by training a linear classifier with random doublets
at varying dimension $n$, number of points $p$, and overlap $\rho$.
Equation (\ref{eq:solution_k2}) matches perfectly as expected.
Equation (\ref{eq:capacity_k2}) is surprisingly precise
even at very small sizes; deviations are less than $1\%$
already for $n=5$.

\section{Polytopes (multiplets, generic $k$)}

Let us now move to the general case where the data
are structured in multiplets of $k$ points.
We consider dichotomies of $k$ sets of points
$\xi^{\mu}=\{\xi^{\mu}_1,\ldots,\xi^{\mu}_p \}$, with $\mu=1,\ldots,k$.
The $i$th multiplet is the set $\xi_i=\{\xi_i^1,\ldots,\xi_i^k\}$.
A dichotomy $\phi$ is admissible if the images of all $k$ partner points
in each multiplet are equal:
$\phi\left(\xi^{\mu}_i\right)=\phi\left(\xi^{\nu}_i\right)$ for all
$\mu,\nu=1,\ldots,k$, separately for all $i=1,\ldots,p$.
For clarity, we denote the number of admissible dichotomies 
by $C_{n,p}^{(k)}$,
as shown in Fig.~\ref{fig:sketch}.

A recursion relation for $C_{n,p}^{(k)}$ can be obtained by 
carefully extending the method used for the doublet case.
At the $(p+1)$th step, we consider the multiplet $\xi_{p+1}$,
composed of the $k$ points $\xi^1_{p+1},\ldots,\xi^k_{p+1}$.
Let us exclude momentarily the point $\xi_{p+1}^1$, and
suppose we know how to apply Cover's method to
the set of $k-1$ points 
\begin{equation}
\bar\xi_{p+1}=\left\{\xi^2_{p+1},\ldots,\xi^k_{p+1}\right\}
\subset\xi_{p+1}.
\end{equation}
This would give an expression, let us call it
\begin{equation}
Q^{k-1} (C_{n,p}^{(k)},C_{n-1,p}^{(k)},\ldots,C_{n-k+1,p}^{(k)}).
\end{equation}
The fact that $Q^{k-1}$ is a function of $C_{n-l,p}^{(k)}$
with $l=0,\ldots,k-1$ will be clear in the following.
Intuitively, the case $k=1$ involves only $l=0$ and $l=1$,
the case $k=2$ adds $l=2$
because it uses the expression for $k=1$ in $n-1$ dimensions,
and the same pattern repeats inductively up to $k-1$ points.

The quantity $Q^{k-1}$ represents
the number of dichotomies of the set
$\xi_1\cup\xi_2\cup\cdots\cup\xi_p\cup\bar\xi_{p+1}$
that are admissible on the first $p$ multiplets
[meaning that
$\phi(\xi_i^\mu)=\phi(\xi_i^\nu)$ for all $\mu,\nu=1,\ldots,k$
and all $i=1,\ldots,p$]
and admissible on the $k-1$ points in $\bar\xi_{p+1}$
[meaning that $\phi(\xi_{p+1}^\mu)=\phi(\xi_{p+1}^\nu)$
for all $\mu,\nu=2,\ldots,k$].
A number $R_{n,p}^{k-1}$ of these dichotomies are realizable by
a hyperplane passing through the excluded point $\xi_{p+1}^1$,
and are therefore all admissible.
Of the remaining $Q^{k-1}(\ldots) - R_{n,p}^{k-1}$ ones,
a fraction $\tilde\Psi_k$ assign the same value
to $\xi_{p+1}^1$ and to the points in $\bar\xi_{p+1}$,
and are therefore admissible on the whole multiplet $\xi_{p+1}$.
Therefore,
\begin{equation}
\label{eq:CkQR}
C_{n,p+1}^{(k)} = \tilde\Psi_k \left[ Q^{k-1}(\ldots) - R_{n,p}^{k-1} \right]
+ R_{n,p}^{k-1}.
\end{equation}

While $\Psi_2$ was a probability (over all possible hyperplanes),
$\tilde\Psi_k$ is a conditional probability, namely
the probability that a uniform vector $w$ on the sphere $S^{n-1}$
does not separate the multiplet $\xi_{p+1}$, conditioned on the event that
$w$ does not separate the set $\bar\xi_{p+1}$:
\begin{equation}\label{eq:psitilde}
\tilde\Psi_k=\frac{
\int_{S^{n-1}} \mathrm{d}w \prod_{\mu,\nu=1}^k 
\theta\left(w\cdot\xi_{p+1}^\mu w\cdot\xi_{p+1}^\nu\right)
}{
\int_{S^{n-1}} \mathrm{d}w \prod_{\mu,\nu=2}^k 
\theta\left(w\cdot\xi_{p+1}^\mu w\cdot\xi_{p+1}^\nu\right)
}.
\end{equation}
The dependence of $\tilde\Psi_k$ on the relative
positions of the points is discussed in the Appendix,
where it is shown that
(i) the calculation of $\tilde\Psi_k$ can be reduced from $n$-dimensional
to $k$-dimensional integrals, and
(ii) $\tilde\Psi_k$ depends on $n$ only through the 
$k(k-1)/2$ overlaps $\rho_{\mu\nu}$ between the points
in a multiplet, which we fix for all multiplets:
\begin{equation}
\rho_{\mu\nu}=\frac{1}{n} \xi_i^\mu \cdot \xi_i^\nu, \quad i=1,\ldots,p; \quad \mu,\nu=1,\ldots,k.
\end{equation}
This property allows us to treat $\tilde\Psi_k$ as a constant
in the recursions, thus simplifying the computations.
Note that, since it is a conditional probability, $\tilde\Psi$
can be written as a ratio of probabilities:
\begin{equation}
\label{eq:psi_tildepsi}
\tilde{\Psi}_k \left(\{ \rho_{\mu\nu}\}_{\mu,\nu=1,\ldots,k} \right) = \frac{\Psi_k 
\left(\{\rho_{\mu\nu}\}_{\mu,\nu=1,\ldots,k}\right)}{\Psi_{k-1}\left(\{\rho_{\mu\nu}\}_{\mu,\nu=2,\ldots,k}\right)},
\end{equation}
where $\Psi_k$ depends on $k(k-1)/2$ overlaps between $k$ points, and denotes the fraction
of hyperplanes not separating the $k$ points.
This definition, together with the identity $\Psi_1=1$, implies that the geometric quantity computed above for $k=2$
is $\Psi_2(\rho)=\tilde\Psi_2(\rho)$.

The number $R_{n,p}^{k-1}$ can be obtained by applying again
Cover's method with respect to the set $\bar\xi_{p+1}$
this time in $n-1$ dimensions because the hyperplane
is constrained to pass through $\xi_{p+1}^1$.
Hence
\begin{equation}
\label{eq:Rnpk}
R_{n,p}^{k-1}=Q^{k-1}(C_{n-1,p}^{(k)},C_{n-2,p}^{(k)},\ldots,C_{n-k,p}^{(k)}).
\end{equation}
Finally, from Eqs.~(\ref{eq:CkQR}) and (\ref{eq:Rnpk}), the recursion for $C_{n,p}^{(k)}$ is
\begin{equation}
C_{n,p+1}^{(k)}=Q^k\left(C_{n,p}^{(k)},C_{n-1,p}^{(k)},\ldots,C_{n-k,p}^{(k)}\right),
\end{equation}
where the functions $Q^k$ (having $k+1$ arguments) satisfy the 
recursive functional relation
\begin{equation}
\label{eq:Qrecursion}
\begin{split}
Q^k\left(x_n,\ldots,x_{n-k}\right)=
\tilde\Psi_k &Q^{k-1}\left(x_n,\ldots,x_{n-k+1}\right)\\
+\left(1-\tilde\Psi_k\right) &Q^{k-1}\left(x_{n-1},\ldots,x_{n-k}\right),
\end{split}
\end{equation}
with the boundary 
$Q^1\left(x_n,x_{n-1}\right)=x_n + x_{n-1}$
given by the form of Eq.~(\ref{eq:cover_recursion}) for a single point.
The recursion in $k$ can be solved, thus yielding again a recursion 
for $C_{n,p+1}^{(k)}$ in $n$ and $p$ only.
Let us call $\theta_k(l)$ the coefficients in the solved recursion:
\begin{equation}
\label{eq:def_theta}
C_{n,p+1}^{(k)}=\sum_{l=0}^k \theta_k(l) C_{n-l,p}^{(k)}.
\end{equation}
Equation (\ref{eq:Qrecursion}) then becomes
\begin{equation}
\label{eq:theta_recursion}
\theta_k(l)=\tilde\Psi_k \theta_{k-1}(l) + \left(1- \tilde\Psi_k \right) \theta_{k-1}(l-1),
\end{equation}
with boundaries $\theta_1(0)=\theta_1(1)=1$ and $\theta_k(-1)=\theta_{k}(k+1)=0$.
For instance, setting $k=2$ in Eqs.~(\ref{eq:def_theta}) and (\ref{eq:theta_recursion})
recovers the recursion for doublets, Eq.~(\ref{eq:the_recursion_k2}), as expected.
For $k=3$ one obtains
\begin{equation}
\label{eq:the_recursion_k3}
\begin{split}
C_{n,p+1}^{(3)} =
&\tilde\Psi_3 \Psi_2 C_{n,p}^{(3)} +
\left[ \tilde\Psi_3 + \Psi_2\left(1-\tilde\Psi_3\right) \right] C_{n-1,p}^{(3)}\\
&+ \left[ \tilde\Psi_3\left(1-\Psi_2\right)+\left(1-\tilde\Psi_3\right) \right] C_{n-2,p}^{(3)}\\
&+ \left(1-\tilde\Psi_3\right)\left(1-\Psi_2\right) C_{n-3,p}^{(3)}.
\end{split}
\end{equation}

\begin{figure}
\includegraphics[scale=1.1]{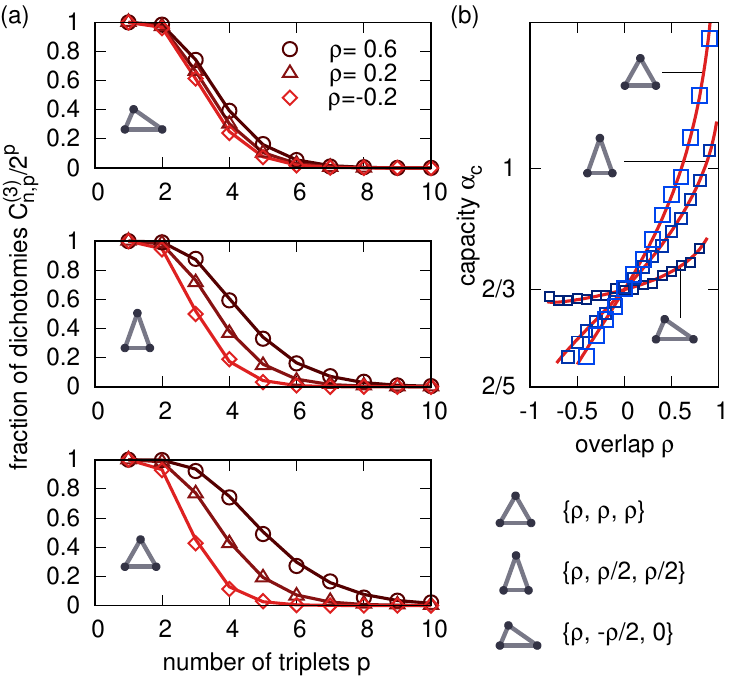}
\caption{
Theory (solid lines) versus numerical results (symbols)
for $k=3$.
(a) Fraction of admissible dichotomies as a function of
the number of triplets $p$ for triplets with overlaps
$\{\rho,\rho,\rho\}$ (bottom),
$\{\rho,\rho/2,\rho/2\}$ (middle),
$\{\rho,-\rho/2,0\}$ (top).
Circles, triangles, and rotated squares correspond to three different values of $\rho$.
The theoretical curves are obtained by solving numerically the recursion,
Eq.~(\ref{eq:the_recursion_k3}).
(b) Capacity as a function of $\rho$ [Eq.~(\ref{eq:capacity_k})] for the same three geometries
(the range of $\rho$ is restricted by the spherical constraint).
}
\label{fig:k3}
\end{figure}

In the process of deriving the foregoing recursion relations
we considered the points $\xi_{p+1}^\mu$ in a particular order,
therefore explicitly breaking
invariance under permutations within the multiplets.
We restore the invariance \emph{a posteriori}, by prescribing that all
$\tilde\Psi_l$ (with $l\leq k$) be symmetrized with respect to all $k(k-1)/2$ overlaps.
For instance, when $k=3$, the $\Psi_2=\tilde\Psi_2$ appearing in Eq.~(\ref{eq:the_recursion_k3})
is to be intended as $[ \Psi_2(\rho_{12}) + \Psi_2(\rho_{13}) + \Psi_2(\rho_{23})]/3$.
The goodness of this prescription is substantiated by
the numerical results shown in Fig.~\ref{fig:k3};
see also the limit case (ii) in the Discussion below.

The solution for $C_{n,p}$ (with the appropriate boundary conditions) 
can be obtained, for instance via generating functions,
but we do not give it here.
Instead, we focus on the capacity, which
can be computed by the same approximate method used for $k=2$
[Eqs.~(\ref{eq:imean}) and (\ref{eq:capacity_k2})]:
\begin{equation}
\label{eq:capacity_moments}
\alpha_\mathrm{c}=\frac{\sum_{l=0}^k \theta_k(l)}{\sum_{l=0}^k l\theta_k(l)}
= \frac{\lambda_0(k)}{\lambda_1(k)},
\end{equation}
where we have defined the moments
\begin{equation}
\lambda_m(k)=\sum_{l=0}^k l^m \theta_k(l).
\end{equation}
Summing Eq.~(\ref{eq:theta_recursion}) over $l$
shows that $\lambda_0(k)=\lambda_0(k-1)$
and therefore $\lambda_0(k)=\lambda_0(1)=2$.
By multiplying Eq.~(\ref{eq:theta_recursion}) by $l$
and summing over $l$, one obtains
$\lambda_1(k)=\lambda_1(k-1)+ (1-\tilde\Psi_k)\lambda_0(k-1)$.
The boundary condition $\lambda_1(1)=1$ then fixes the solution
\begin{equation}
\label{eq:lambda1}
\lambda_1(k)=2k-1-2\sum_{l=2}^k \tilde\Psi_l.
\end{equation}
Finally, substituting $\lambda_0(k)$ and $\lambda_1(k)$ into Eq.~(\ref{eq:capacity_moments})
yields a remarkably simple formula for the capacity:
\begin{equation}
\label{eq:capacity_k}
\alpha_\mathrm{c}=\left(
k-\frac{1}{2}-\sum_{l=2}^k \tilde\Psi_l
\right)^{-1}.
\end{equation}

Figure~\ref{fig:k3} compares our theory with numerical computations
in the case of triplets ($k=3$), for triangles with three, two, and no sides of the same length.
The agreement is excellent.
The function $\tilde\Psi_3$ is a double integral (given in the Appendix),
which we evaluate numerically.

\section{Discussion}
Our extension of Cover's combinatorial technique
to structured data allows to obtain closed
expressions of $C_{n,p}^{(k)}$ at finite $n$ and $p$, for any $k$
[we have written explicitly the result for $k=2$ in Eq.~(\ref{eq:solution_k2})].
Beside this, our main result is Eq.~(\ref{eq:capacity_k}),
which expresses the capacity as a simple function of
the quantities $\tilde\Psi_l$.
Regarding these quantities, the merit of our method is twofold:
first, the $\tilde\Psi_l$'s are revealed to be the only
relevant parameters characterizing the linear separability of the multiplets;
second, they have a very simple geometric interpretation
in terms of probabilities.

We mention three simple limit cases of Eq.~(\ref{eq:capacity_k}).
(i) If all the points in each multiplet coincide, then
$\tilde\Psi_l=1$ for all $l=1,\ldots,k$ and we recover the
single-point classic result $\alpha_\mathrm{c}=2$.
(ii) When $k=3$ and two points of a triplet coincide the overlaps are
$\{\rho,\rho,1\}$.  Symmetrizing $\tilde\Psi_3(\rho,\rho,1)$ gives
$\Psi_3(\rho,\rho,1) \left[2/\Psi_2(\rho)+1/\Psi_2(1)\right]/3$ where
$\Psi_3(\rho,\rho,1)$ is the fraction of hyperplanes not separating
the three points.  
Clearly
$\Psi_3(\rho,\rho,1)=\Psi_2(\rho)$, and one recovers
Eq.~(\ref{eq:capacity_k2}) for $k=2$ as expected.
(iii) If $\Psi_2=0$ and $\tilde\Psi_l=0$ for all $l=3,\ldots,k$
Eq.~(\ref{eq:capacity_k}) gives $\alpha_\mathrm{c}=2/(2k-1)$.  This
prediction matches that obtained in \cite{ChungLeeSompolinsky:2018}
for $(k-1)$-dimensional linear manifolds.  However, this turns out to be
an unphysical limit in our framework, since $\tilde\Psi_l$ cannot be
all vanishing.  For instance, for $k=3$, equilateral triplets with
overlaps $\{\rho,\rho,\rho\}$ lie on a linear manifold passing through
the origin when $\rho$ takes its minimum value $\rho_\triangle=-1/2$.
The same happens for isosceles triplets $\{\rho,\rho/2,\rho/2\}$ at
$\rho_\vartriangle=1-\sqrt{3}.$ Interestingly, the capacity evaluated
at the respective minimum $\rho$ is $\alpha_\mathrm{c}\approx 0.46154$
for both geometries, to be compared to the value
$\alpha_\mathrm{c} = 2/5$ found for two-dimensional linear manifolds.

Another interesting, albeit less elementary, limit case would be
$k\to\infty$, taken in such a way that the points generate a sphere of
radius $\kappa$; then Eq.~(\ref{eq:capacity_k}) should reproduce the
well-known capacity with margin $\kappa$ \cite{Gardner:1987},
which has never been obtained by combinatorial methods
\cite{ChungLeeSompolinsky:2018,EngelVanDenBroeck:BOOK}.

Other applications and extensions of the theory appear possible.
First, the capacity is written in Eq.~(\ref{eq:capacity_moments}) as a
combination of the zeroth and first moments, but higher-order moments
can be computed similarly and give access to other useful quantities.
For instance, the second moment is related to the width of the
crossover region separating the regimes where $c_{n,p}\approx 1,0$
respectively.
Second, it would be interesting to express our results for general
(non-linear) separating surfaces, in the same spirit of Cover's
original work, and in view of useful applications.

\acknowledgments
We would like to dedicate this work to the memory of Bruno Bassetti.
P.R.~acknowledges funding by the European Union through the H2020 - MCIF Grant No.~766442.

\appendix*

\section{Computation of $\Psi_k$}

\paragraph{Computation of $\Psi_2(\rho)$.}

The fraction of hyperplanes assigning the same value to two points $\xi$ and $\bar\xi$ is given by: 
\begin{equation}\label{eq:def_Psi}
\Psi_2 = \frac{2}{\mathcal N}\int \! \mathrm{d}^nx \; \delta\left(\norm{x}^2-1\right) \theta\left(x\cdot \xi\right)\theta\left(x\cdot \bar\xi\right)
\end{equation}
The normalization factor is
\begin{equation}\label{eq:normalization}
\mathcal N = \int \! \mathrm{d}^nx \; \delta\left(\norm{x}^2-1\right) = \Omega_n/2,
\end{equation}
where $\Omega_n$ is the solid angle in $n$ dimensions.
Gram-Schmidt (GS) orthonormalization of $\xi$ and $\bar\xi$ yields
\begin{equation}\label{eq:GS}
e_1 = \xi, \quad e_2 = \frac{\bar\xi - \rho\, \xi}{\sqrt{1-\rho^2}},
\end{equation}
where $\rho = \xi \cdot \bar\xi/n$ is the overlap between the two points.
Inverting Eq.~(\ref{eq:GS}) gives
\begin{equation}
\xi = e_1\,, \quad \bar\xi = \rho\, e_1 + \sqrt{1-\rho^2}\, e_2.
\end{equation}
Having orthonormalized the points allows to safely exploit the $(n-2)$-dimensional spherical symmetry of the integral in the space orthogonal to $\xi_1$ and $\xi_2$, and to reduce it to an integral over the two-dimensional solid angle:
\begin{equation}
\Psi_2 = \int \frac{\mathrm{d}\Omega_2}{\pi} \theta \left(\cos \phi\right) \theta \left(\rho\cos \phi +\sqrt{1-\rho^2} \sin \phi \right),
\end{equation}
which evaluates to the result in Eq.~(\ref{eq:Psi2}),
and shows that $\Psi_2=\Psi_2(\rho)$.

\paragraph{Computation of $\Psi_3 (\rho_{12},\rho_{13},\rho_{23})$.}

Eq.~(\ref{eq:psi_tildepsi}) expresses the conditional probability $\tilde\Psi_k$
in terms of the probabilities $\Psi_k$.
$\Psi_k$ is
defined as the fraction of hyperplanes assigning the same value to the $k$ points $\xi_1,\xi_2,\dots,\xi_k$:
\begin{equation}\label{eq:Psik}
\Psi_k =  \frac{2}{\mathcal{N}} \int \! \mathrm{d}^nx \; \delta\left(\norm{x}^2-1\right)  \prod_{\mu =1}^k \theta (x\cdot \xi^{\mu}),
\end{equation}
with $\mathcal N$ given by Eq.~(\ref{eq:normalization}).
For $k=3$, the Gram-Schmidt procedure gives:
\begin{equation*}
e_1 = \xi^1\,, \quad e_2 = \frac{\xi^2 - \rho_{12} \xi^1}{\sqrt{1-\rho_{12}^2}}\,, \quad e_3 = \frac{\xi^3 - \rho_{13} e_1 - g e_2}{\sqrt{1-\rho_{13}^2- g^2}}\,,
\end{equation*} 
where $\rho_{\mu\nu}=\xi^\mu \cdot \xi^\nu/n$ are the overlaps, and $g=(\rho_{23}-\rho_{12} \rho_{13})/\sqrt{1-\rho_{12}^2}$. 
Again, thanks to the spherical symmetry in the space orthogonal to the $\xi^\mu$'s 
the result is an integral over the $3$-dimensional solid angle:
\begin{equation}\label{eq:psi3}
\begin{split}
\Psi_3 = &\frac{\Gamma \left(\frac{3}{2}\right)}{\pi^{\frac{3}{2}}} \int \mathrm{d} \Omega_3	\,  \theta \left(\rho_{12} x_1+ \sqrt{1-\rho_{12}^2} x_2 \right) \\ & \theta \left(x_1 \right) \theta \left(\rho_{13} x_1+ g x_2+ \sqrt{1-\rho_{13}^2-g^2} x_3 \right)\,,
\end{split}
\end{equation}
where the measure $\mathrm{d}\Omega_3$ can be expressed via the angles $\phi_1$ and $\phi_2$, and
$x_1 = \sin \phi_1 \cos \phi_2$, $x_2 = \sin \phi_1 \sin \phi_2$ and $x_3 = \cos \phi_1$.
As above, this computation shows that $\Psi_3=\Psi_3 (\rho_{12},\rho_{13},\rho_{23})$.
The results presented in Fig.~\ref{fig:k3} have been obtained by integrating numerically Eq.~(\ref{eq:psi3}).

The procedure for $k=2,3$ can be extended to $k > 3$. The final result has the following structure:
\begin{equation*}
\Psi_k (\{\rho_{\mu\nu}\}) = \frac{\Gamma \left(\frac{k}{2}\right)}{\pi^{\frac{k}{2}}} \int \mathrm{d} \Omega_k (\phi_1,\phi_2,\dots,\phi_k) \prod_{\alpha =1}^k \theta \left(v_{\alpha} (\phi) \right)\,, 
\end{equation*} 
where the functions $v_{\alpha}$ appearing in the $\theta$'s can be systematically derived in a similar way from the GS procedure.
This shows that $\tilde\Psi_k$, related to $\Psi_k$ by Eq.~(\ref{eq:psi_tildepsi}),
depends in general on the $\xi^\mu$'s only through the overlaps $\rho_{\mu\nu}$,
and it can be written in terms of $k$-dimensional integrals.

\bibliography{biblio}

\end{document}